\documentclass{amsart}
\usepackage{amsfonts}

\usepackage{graphicx}

\input{tcilatex}
\input tcilatex

\begin{document}
\title[combinatorial population dynamics]{Some combinatorial aspects of discrete non-linear population dynamics}
\author{Nicolas Grosjean, Thierry Huillet}
\address{Laboratoire de Physique Th\'{e}orique et Mod\'{e}lisation CNRS-UMR 8089 et
Universit\'{e} de Cergy-Pontoise, 2 Avenue Adolphe Chauvin, 95302,
Cergy-Pontoise, FRANCE\\
E-mail(s): Nicolas.Grosjean@u-cergy.fr and Thierry.Huillet@u-cergy.fr}
\maketitle

\begin{abstract}
Motivated by issues arising in population dynamics, we consider the problem
of iterating a given analytic function a number of times. We use the
celebrated technique known as Carleman linearization that turns (for a
certain class of functions) this problem into simply taking the power of a
real number. We expand this method, showing in particular that it can be
used for population models with immigration, and we also apply it to the
famous logistic map. We also are able to give a number of results for the
invariant density of this map, some being related to the Carleman
linearization.\newline

\textbf{Keywords:} non-linear population dynamics without and with
immigration, Carleman transfer matrix, logistic map.
\end{abstract}

\section{Introduction}

We consider simple $1-$dimensional discrete-time dynamics: $x_{n+1}=\phi
\left( x_{n}\right) $, $x_{0}=x,$\ with the evolution mechanism $\phi \left(
\cdot \right) $\ being an analytic function. Our main interest in these
problems arises from population dynamics models describing the temporal
evolution of some population with size\textbf{\ }$x_{n}\geq 0$\textbf{.} We
first assume $\phi \left( 0\right) =0$ (no immigration). Such non-linear
models are amenable to a Carleman linearization giving $x_{n}$ from the
initial condition $x$ in terms of the $n-$th power of some upper-triangular
infinite-dimensional transfer matrix which can be diagonalized.
Equivalently, $\phi $ is $h$-conjugate to the linear map $\lambda x$ for
some Carleman function $h$, would $\lambda =\phi ^{\prime }\left( 0\right)
\neq \left\{ -1,0,1\right\} $. The coefficients of $h$, as a power series in 
$x$, are obtained from the left eigenvector of $P$ with the eigenvalue $%
\lambda $. The Carleman linearization technique goes back the 60' (\cite
{Erdos}, \cite{Jab}, \cite{Ko}, \cite{GK}). When $\lambda =1$ (the critical
case), we give the linear Carleman representation of $x_{n}$, using a
`Jordanization' technique. Special such models arising in population
dynamics are defined and investigated.\textbf{\ }We next consider the
problem of computing the invariant density (and its support)\textbf{\ }of
the dynamics in a chaotic population model regime, including quadratic maps.
The study of the invariant measures of quadratic and related maps has a very
long story starting in the 70' (\cite{Buni}, \cite{Ruelle}, \cite{Bowen}, 
\cite{Jako}, \cite{Collet}). We show that in some special cases, the $h$%
-conjugate representation of $\phi $ is useful for that purpose. We
illustrate our point of view on the celebrated logistic population model $%
\phi \left( x\right) =rx\left( 1-x\right) $. Next we consider $\phi
_{0}\left( x\right) =c+\phi \left( x\right) $,\textbf{\ }modelling some
population dynamics with immigration\textbf{\ }$c>0$\textbf{.} In the
presence of a fixed point for $\phi _{0}$, such models are also Carleman
linearizable; equivalently, $\phi _{0}$ is shown to be $g$-conjugate now to
an affine map for some explicit Carleman function $g$. As an illustration%
\textbf{,} we finally deal with the logistic population model with
immigration. We develop its intimate relation to a family of companion
logistic population models without immigration, the former being obtained
from the latter through a suitable affine transformation. We exploit this
deep connection to determine under which condition the logistic model with
immigration is chaotic or not and, using this observation, we compute in
some cases its invariant density.

The precise organization of the paper is as follows:

In Section 2, we recall and develop the Carleman transfer matrix
linearization technique, including:

- its link with a conjugate representation of the map $\phi $\ in the case $%
\phi \left( 0\right) =0$ and $\phi ^{\prime }\left( 0\right) \neq \left\{
-1,0,1\right\} .$

- the application of this scheme to the specific critical case $\phi
^{\prime }\left( 0\right) =1$, leading to a method akin to the Jordanization
of a matrix.

- the consequences of this construction in terms of the invariant measure of
the dynamical system.

- the application of this general setting to a class of specific population
evolution models.

In Section 3, using the above tools, we focus on the one-parameter logistic
population model. Specifically, we characterize the loci and the types of
the divergence of its invariant measure and we give a way to compute the
disconnected components of its support for some parameter $r$\ range. For
some values of the parameter $r$, we show how to compute explicitly the
invariant density of the system.

In Section 4, expanding the tools introduced in Section 2 to include maps
obeying $\phi _{0}\left( 0\right) \neq 0$, we study the effect of adding
immigration to population dynamics models, by relating it to a an affine
conjugate equivalent of the new mechanism with immigration $\phi _{0}\left(
0\right) >0$. Once again, we apply these results to the logistic map. Our
main result on this point is summarized in Figure $1$ showing the values of
the parameters $\left( r,c\right) $\ for which this topological conjugation
is admissible. As a consequence, the chaoticity of the logistic model with
immigration is revealed by the one of the corresponding model without
immigration.

\section{Carleman matrix in the triangular case $\alpha _{0}=0$}

With $\alpha _{k}$, $k\geq 1$, real numbers, let $\phi \left( x\right)
=\sum_{k\geq 1}\alpha _{k}x^{k}$, $\alpha _{1}\neq 0$, be some smooth power
series defined (convergent) in some neighborhood $x_{c}^{-}<x<x_{c}^{+}$, of
the origin where $-\infty \leq x_{c}^{-}<0<x_{c}^{+}\leq \infty $\footnote{%
We also assume that $\phi $ is absolutely convergent with radius of
convergence $0<r_{c}\leq \min \left( -x_{c}^{-},x_{c}^{+}\right) $}. We
avoid the trivial linear case $\phi \left( x\right) =\alpha _{1}x$. We shall
let $I_{c}=\left( x_{c}^{-},x_{c}^{+}\right) $ be the interval of
convergence. Note $\alpha _{0}=0$. Consider the dynamical system 
\begin{equation}
x_{n+1}=\phi \left( x_{n}\right) \text{, }x_{0}=x.  \label{e1}
\end{equation}
Define the infinite-dimensional (Carleman) upper-triangular matrix 
\begin{equation}
P\left( k,k^{\prime }\right) =\left[ x^{k^{\prime }}\right] \phi \left(
x\right) ^{k}\text{, }k^{\prime }\geq k\geq 1.  \label{e2}
\end{equation}
By Fa\`{a} di Bruno formula (see e.g. \cite{Comtet}, Tome $1$, p. $148$),
with $\widehat{B}_{k,l}\left( \alpha _{1},\alpha _{2},...\right) $ the
(ordinary) Bell polynomials in the coefficients $\alpha _{k}:=\left[
x^{k}\right] \phi \left( x\right) $ of $\phi \left( x\right) $, 
\begin{equation}
P\left( k,k^{\prime }\right) =\widehat{B}_{k^{\prime },k}\left( \alpha
_{1},\alpha _{2},...,\alpha _{k^{\prime }-k+1}\right)
=k!\sum_{c_{l}}^{**}\prod_{l=1}^{k^{\prime }-k+1}\frac{\alpha _{l}^{c_{l}}}{%
c_{l}!},  \label{e3}
\end{equation}
where the last double-star summation runs over the integers $c_{l}\geq 0$
such that $\sum_{l=1}^{k^{\prime }-k+1}c_{l}=k$ and $\sum_{l=1}^{k^{\prime
}-k+1}lc_{l}=k^{\prime }$ (there are $p_{k,k^{\prime }}$ terms in this sum,
the number of partitions of $k^{\prime }$ into $k$ summands). In particular $%
P\left( k,k\right) =\alpha _{1}^{k}$ and $P\left( k,k+1\right) =k\alpha
_{2}\alpha _{1}^{k-1}$. $P$ is called the Carleman\footnote{%
Carleman matrices are easily seen to be the transpose of Bell matrices.}
matrix of $\phi $. If for example, $\phi \left( x\right) =x-x^{2}$, $P\left(
k,k^{\prime }\right) =\widehat{B}_{k^{\prime },k}\left( z,-1,0,...\right)
\mid _{z=1},$ the Hermite polynomials evaluated at $z=1$. We conclude (see 
\cite{Jab}, \cite{Rab}, \cite{Ber}, \cite{Ko} and \cite{GK}):\newline

\textbf{Proposition:} With $\mathbf{e}_{1}^{\prime }=\left( 1,0,0,...\right) 
$ and $\mathbf{x}^{\prime }=\left( x,x^{2},...\right) $\footnote{%
Throughout, a boldface variable, say $\mathbf{x}$, will represent a
column-vector and its transpose, say $\mathbf{x}^{\prime }$, will be a
row-vector.}, 
\begin{equation}
x_{n}=\mathbf{e}_{1}^{\prime }P^{n}\mathbf{x}  \label{e4}
\end{equation}
where $P$ is an upper-triangular `transfer' matrix with $P\left( k,k\right)
=\alpha _{1}^{k}=:\lambda ^{k}$, $k\geq 1$ (the eigenvalues of $P$).

From (\ref{e1}), $x_{n}$ is also $x_{n}=\phi _{n}\left( x\right) $ where $%
\phi _{n}$ is the $n-$th iterate of $\phi $ by composition and so (\ref{e4})
is an alternative linear representation of $x_{n}$. Note 
\begin{equation*}
\sum_{n\geq 0}\lambda ^{n}x_{n}=\mathbf{e}_{1}^{\prime }\left( I-\lambda
P\right) ^{-1}\mathbf{x,}
\end{equation*}
involving the resolvent of $P$.\newline

\emph{Remark:} We have 
\begin{eqnarray*}
\frac{1}{1-u\phi \left( x\right) } &=&1+\sum_{k\geq 1}u^{k}\phi \left(
x\right) ^{k}=1+\sum_{k\geq 1}u^{k}\sum_{k^{\prime }\geq k}x^{k^{\prime
}}\left[ x^{k^{\prime }}\right] \phi \left( x\right) ^{k} \\
&=&1+\sum_{k\geq 1}u^{k}\sum_{k^{\prime }\geq k}x^{k^{\prime }}P\left(
k,k^{\prime }\right) =1+\sum_{k^{\prime }\geq 1}x^{k^{\prime
}}\sum_{k=1}^{k^{\prime }}u^{k}\left[ x^{k^{\prime }}\right] \phi \left(
x\right) ^{k}
\end{eqnarray*}
\begin{equation*}
\sum_{k=1}^{k^{\prime }}u^{k}P\left( k,k^{\prime }\right) =\left[
x^{k^{\prime }}\right] \frac{1}{1-u\phi \left( x\right) }.
\end{equation*}
With $\phi _{n}\left( x\right) =x_{n}$, the $n-$th iterate of $\phi $\ and $%
\mathbf{u}^{\prime }:=\left( u,u^{2},...\right) $%
\begin{equation*}
\frac{1}{1-u\phi _{n}\left( x\right) }=1+\sum_{k\geq 1}u^{k}\sum_{k^{\prime
}\geq k}x^{k^{\prime }}P^{n}\left( k,k^{\prime }\right) =1+\sum_{k^{\prime
}\geq 1}x^{k^{\prime }}\sum_{k=1}^{k^{\prime }}u^{k}P^{n}\left( k,k^{\prime
}\right) =1+\mathbf{u}^{\prime }P^{n}\mathbf{x.}
\end{equation*}
Taking the derivative with respect to $u$\ at $u=0$\ gives $\phi _{n}\left(
x\right) =x_{n}=\mathbf{e}_{1}^{\prime }P^{n}\mathbf{x.}$ Taking the $k$-th
derivative with respect to $u$\ at $u=0$\ gives $x_{n}^{k}=\mathbf{e}%
_{k}^{\prime }P^{n}\mathbf{x.}$ We conclude:\newline

\textbf{Proposition:} If $\psi \left( x\right) =\sum_{k\geq 1}\psi _{k}x^{k}$%
\ is some smooth observable, defining $\mathbf{\psi }^{\prime }=\left( \psi
_{1},\psi _{2},...\right) $, therefore 
\begin{equation}
\psi \left( x_{n}\right) =\sum_{k}\psi _{k}\mathbf{e}_{k}^{\prime }P^{n}%
\mathbf{x}=\mathbf{\psi }^{\prime }P^{n}\mathbf{\mathbf{x}.}  \label{e4a}
\end{equation}
This generalizes (\ref{e4}).

By Cauchy formula, whenever $\phi $\ is defined on the unit circle, we also
have the Fourier representation 
\begin{equation*}
P\left( k,k^{\prime }\right) =\frac{1}{2\pi }\int_{0}^{2\pi }e^{ik^{\prime
}\theta }\phi \left( e^{-i\theta }\right) ^{k}d\theta .
\end{equation*}

Chaos for (\ref{e1}) is sometimes characterized by the positivity of its
Lyapounov exponent defined by 
\begin{equation*}
\lambda \left( x\right) =\lim_{N\rightarrow \infty }\frac{1}{N}%
\sum_{n=0}^{N-1}\log \left| \phi ^{\prime }\left( x_{n}\right) \right| ,
\end{equation*}
for almost all $x$. Considering the sensitivity to the initial condition
problem, we have $J_{n+1}:=dx_{n+1}/dx=\phi ^{\prime }\left( x_{n}\right)
dx_{n}/dx=\phi ^{\prime }\left( x_{n}\right) J_{n}$. Therefore $\left|
J_{N}\right| =\left( \prod_{n=0}^{N-1}\left| \phi ^{\prime }\left(
x_{n}\right) \right| \right) $\ and $\lambda _{N}\left( x\right) :=\frac{1}{N%
}\log \left| J_{N}\right| \rightarrow \lambda \left( x\right) $. Letting $%
\mathbf{x}^{\prime }=\left( x,x^{2},...\right) ,$\ $\overline{\mathbf{x}}%
^{\prime }:=\left( 1,x,x^{2},...\right) $ and $D:=$diag$\left(
1,2,3,...\right) $ so that $d\mathbf{x}/dx=D\overline{\mathbf{x}}$, we
observe 
\begin{equation*}
J_{N}:=dx_{N}/dx=\mathbf{e}_{1}^{\prime }P^{N}D\overline{\mathbf{x}}.
\end{equation*}

\subsection{The case $\left| \lambda \right| \neq 1$}

Suppose $\lambda :=\alpha _{1}\neq \pm 1$ and let $\mathbf{v}_{k}^{\prime
}P=\lambda ^{k}\mathbf{v}_{k}^{\prime }$, define the left-row-eigenvector $%
\mathbf{v}_{k}^{\prime }$ of $P$ associated to the eigenvalue $\lambda
_{k}:=\lambda ^{k}$. Then, with $k^{\prime }>k\geq 1,$%
\begin{equation*}
\mathbf{v}_{k}\left( k^{\prime }\right) =\left( \lambda ^{k}-\lambda
^{k^{\prime }}\right) ^{-1}\sum_{l=1}^{k^{\prime }-1}P\left( l,k^{\prime
}\right) \mathbf{v}_{k}\left( l\right)
\end{equation*}
gives the entries $\mathbf{v}_{k}\left( k^{\prime }\right) $, $k^{\prime
}>k\geq 1$, of $\mathbf{v}_{k}^{\prime }$ by recurrence; and $\mathbf{v}%
_{k}\left( k\right) $ can be left undeterminate. Developing, for $k^{\prime
}>k$, we get\newline

\textbf{Proposition: }Suppose\textbf{\ }$\lambda =\alpha _{1}\neq 1$, then 
\begin{eqnarray*}
\mathbf{v}_{k}\left( k^{\prime }\right) &=&\mathbf{v}_{k}\left( k\right)
\sum_{j=2}^{k^{\prime }-k+1}\sum_{d_{1}+..+d_{j-1}=k^{\prime
}-k}^{*}\prod_{l=1}^{j-1}\frac{P\left(
d_{1}+..+d_{l-1}+k,d_{1}+..+d_{l}+k\right) }{\lambda ^{k}-\lambda
^{d_{1}+..+d_{l}+k}},\text{ or} \\
\mathbf{v}_{k}\left( k^{\prime }\right) &=&\mathbf{v}_{k}\left( k\right)
\sum_{j=2}^{k^{\prime }-k+1}\sum_{k=k_{1}<k_{2}<..<k_{j-1}<k_{j}=k^{\prime
}}\prod_{l=1}^{j-1}\frac{P\left( k_{l},k_{l+1}\right) }{\lambda ^{k}-\lambda
^{k_{l+1}}},
\end{eqnarray*}
where the star-sum in the first identity (involving the integer partition of 
$k^{\prime }-k$) runs over the integers $d_{l}\geq 1$ summing to $k^{\prime
}-k$. \newline

Thus, with $V=\left[ \mathbf{v}_{1},...,\mathbf{v}_{k},...\right] ^{\prime }$%
, $VP=D_{\mathbf{\lambda }}V$ where $V$ is upper-triangular (so here
invertible) and $D_{\mathbf{\lambda }}=$diag$\left( \lambda ,\lambda
^{2},...\right) $, 
\begin{equation}
x_{n}=\mathbf{e}_{1}^{\prime }V^{-1}D_{\mathbf{\lambda }}^{n}V\mathbf{x.}
\label{e5}
\end{equation}
This shows that a general non-linear dynamical system (\ref{e1}) generated
by $\phi $ with $\phi \left( 0\right) =\alpha _{0}=0$ and $\alpha _{1}\neq
\left\{ -1,0,1\right\} $ is in fact a linear infinite-dimensional system
with `transfer matrix' $P$ which can effectively be diagonalized.\newline

\textbf{Proposition:} Defining $h_{k}\left( x\right) =\mathbf{v}_{k}^{\prime
}\mathbf{x}=\sum_{k^{\prime }\geq k}\mathbf{v}_{k}\left( k^{\prime }\right)
x^{k^{\prime }}$, this also means that for all $k\geq 1$%
\begin{equation*}
x_{n}=h_{k}^{-1}\left( \lambda ^{nk}h_{k}\left( x\right) \right) .
\end{equation*}

Proof: Indeed, $\left( P\mathbf{x}\right) _{l}=\phi \left( x\right) ^{l}$
and $\mathbf{v}_{k}^{\prime }P=\lambda ^{k}\mathbf{v}_{k}^{\prime
}\Rightarrow \mathbf{v}_{k}^{\prime }P\mathbf{x}=\lambda ^{k}\mathbf{v}%
_{k}^{\prime }\mathbf{x\Rightarrow }h_{k}\left( \phi \left( x\right) \right)
=\lambda ^{k}h_{k}\left( x\right) $ where $h_{k}\left( x\right) =\mathbf{v}%
_{k}^{\prime }\mathbf{x}=\sum_{k^{\prime }\geq k}\mathbf{v}_{k}\left(
k^{\prime }\right) x^{k^{\prime }}$, for all $k\geq 1.$ Iterating, $%
x_{1}=\phi \left( x\right) =h_{k}^{-1}\left( \lambda ^{k}h_{k}\left(
x\right) \right) $ $\Rightarrow $ $x_{2}=\phi \left( x_{1}\right)
=h_{k}^{-1}\left( \lambda ^{2k}h_{k}\left( x\right) \right) $ $\Rightarrow $ 
$x_{n}=\phi \left( x_{n-1}\right) =h_{k}^{-1}\left( \lambda ^{nk}h_{k}\left(
x\right) \right) .$

When $k=1$, with $\mathbf{v:}=\mathbf{v}_{1}$ and 
\begin{equation}
h\left( x\right) =\mathbf{v}^{\prime }\mathbf{x}=\sum_{k^{\prime }\geq 1}%
\mathbf{v}\left( k^{\prime }\right) x^{k^{\prime }}=:\sum_{k^{\prime }\geq
1}h_{k^{\prime }}x^{k^{\prime }}  \label{e6}
\end{equation}
and $h^{-1}$ the analytic continuation of the inverse function of $h$, both
obeying $h\left( 0\right) =h^{-1}\left( 0\right) =0$, this in particular
means 
\begin{equation}
x_{1}:=\phi \left( x\right) =h^{-1}\left( \lambda h\left( x\right) \right) 
\text{ and }x_{n}=h^{-1}\left( \lambda ^{n}h\left( x\right) \right) .
\label{e7}
\end{equation}
The importance of this formula relies on the fact that, in this way, $x_{n}$
can be evaluated directly, for any initial point $x$, without actually
computing the intermediate values $x_{1},...,x_{n-1}$. But this is at the
expense of the computation of $h$ and $h^{-1}$, which are simple special
functions only in some exceptional situations.\newline

\emph{Remarks:}

$\left( i\right) $In a neighborhood of the origin, the analytic continuation
of the inverse function and the inverse function itself defined on the range
of $h$ indeed coincide. By Lagrange inversion formula, $h^{-1}$, viewed as a
power series, is locally defined in some neighborhood of the origin\footnote{%
If $h$ has a positive convergence radius, so does $h^{-1}$.}, with
coefficients 
\begin{equation*}
g_{k}:=\left[ x^{k}\right] h^{-1}\left( x\right) =\frac{1}{k}\left[
x^{k-1}\right] \left( h\left( x\right) /x\right) ^{-k}.
\end{equation*}
By Fa\`{a} di Bruno formula, $g_{1}=\left( h^{-1}\right) ^{\prime }\left(
0\right) =1/h_{1}$ and for $k\geq 2$%
\begin{equation}
g_{k}=h_{1}^{-k}\sum_{l=1}^{k-1}\left( -1\right) ^{l}\left( k+l-1\right)
_{k-1}\widehat{B}_{k-1,l}\left( \widehat{h}_{1},\widehat{h}_{2},...\right) ,
\label{e7a}
\end{equation}
where $\left( k\right) _{l}=k\left( k-1\right) ...\left( k-l+1\right) $, $%
\widehat{h}_{k}=h_{k+1}/h_{1}$ and $\widehat{B}_{k,l}$ the `ordinary' Bell
polynomials.

$\left( ii\right) $If $0<\lambda <1$ (a subcritical case), (\ref{e7}) says $%
x_{n}\rightarrow 0$ whatever $x$ as $n\rightarrow \infty $. Furthermore, a
first order Taylor development shows that $x_{n}=h^{-1}\left( 0\right)
+\left( h^{-1}\right) ^{\prime }\left( 0\right) \lambda ^{n}h\left( x\right) 
$ with $h^{-1}\left( 0\right) =0$ and $\left( h^{-1}\right) ^{\prime }\left(
0\right) =1/h_{1}$. So $x_{n}$ goes to $0$ geometrically fast at rate $%
\lambda $ and $\lambda ^{-n}x_{n}\rightarrow h\left( x\right) /h_{1}$.%
\newline

\textbf{Proposition:} If we choose $\mathbf{v}_{k}\left( k\right) =\mathbf{v}%
_{1}\left( 1\right) ^{k}$, then $h_{k}\left( x\right) =h_{1}\left( x\right)
^{k}=:h\left( x\right) ^{k}.$

Proof: Consider the ratio $r\left( x\right) =h_{k}\left( x\right) /h\left(
x\right) ^{k}$. It holds that $r\left( \phi \left( x\right) \right) =r\left(
x\right) $ because 
\begin{equation*}
r\left( \phi \left( x\right) \right) =h_{k}\left( \phi \left( x\right)
\right) /h\left( \phi \left( x\right) \right) ^{k}=\frac{\lambda
^{k}h_{k}\left( x\right) }{\lambda ^{k}h_{1}\left( x\right) ^{k}}=r\left(
x\right) \text{.}
\end{equation*}
But then all the $k-$derivatives $r^{\left( k\right) }\left( 0\right) $, $%
k\geq 1$, of $r$ at $x=0$ vanish because $\phi \left( 0\right) =0$ and $\phi
^{\prime }\left( 0\right) =\lambda \neq 1$. Therefore $r$ is constant on $%
x\in I_{c}$. Choosing $\mathbf{v}_{k}\left( k\right) =\mathbf{v}_{1}\left(
1\right) ^{k}$, then $r\left( 0\right) =\lim_{x\rightarrow 0}h_{k}\left(
x\right) /h\left( x\right) ^{k}=\mathbf{v}_{k}\left( k\right) /\mathbf{v}%
_{1}\left( 1\right) ^{k}=1$, so $r\left( x\right) =1.$ One possible way to
achieve this is to assume without loss of generality that $\mathbf{v}%
_{1}\left( 1\right) =h_{1}=1=\mathbf{v}_{k}\left( k\right) .$\newline

\textbf{Corollary:} If $\mathbf{v}_{k}\left( k\right) =\mathbf{v}_{1}\left(
1\right) ^{k}$, the upper-triangular matrices $V$ and $V^{-1}$ in (\ref{e5})
are the Carleman transfer matrices of the power series $h$ and $h^{-1}$ in (%
\ref{e7}); similarly, the matrix $D_{\mathbf{\lambda }}$ is the transfer
matrix of the linear power series $x\rightarrow \lambda x.$\newline

Proof: by definition $V\left( k,k^{\prime }\right) =\mathbf{v}_{k}\left(
k^{\prime }\right) =\left[ x^{k^{\prime }}\right] h_{k}\left( x\right)
=\left[ x^{k^{\prime }}\right] h\left( x\right) ^{k}$ and by Lagrange
inversion formula, $V^{-1}\left( k,k^{\prime }\right) =\left[ x^{k^{\prime
}}\right] h^{-1}\left( x\right) ^{k}.$\newline

\subsection{The critical case $\lambda =\alpha _{1}=1$ and Jordanization of $%
P$}

Let us consider the critical case, $\alpha _{0}=0$ and $\alpha _{1}=1$. The
transfer matrix $P$ has only $1$s on its diagonal, and is subsequently not
(infinite-)diagonalizable. Let us search instead for a family of column
vectors $\mathbf{w}_{1}$, $\mathbf{w}_{2}$, \dots , such that 
\begin{equation*}
P\mathbf{w}_{1}=\mathbf{w}_{1}
\end{equation*}
and, if $k>1$, 
\begin{equation*}
P\mathbf{w}_{k}=\mathbf{w}_{k-1}+\mathbf{w}_{k},
\end{equation*}
which corresponds to the Jordanization of the infinite matrix $P$. A
solution is given by $\mathbf{w}_{1}^{\prime }=(1,0,\dots )$, $\mathbf{w}%
_{k}(k^{\prime })=0$ if $k^{\prime }>k$, $\mathbf{w}_{k}(1)=0$ if\footnote{%
This condition is not necessary, but it simplifies the computation.} $k>1$
and otherwise 
\begin{equation*}
\mathbf{w}_{k}(k^{\prime })=\frac{1}{P(k^{\prime }-1,k^{\prime })}\left( 
\mathbf{w}_{k-1}(k^{\prime }-1)-\sum_{i=k^{\prime }+1}^{k}P(k^{\prime }-1,i)%
\mathbf{w}_{k}(i)\right)
\end{equation*}
With $W$ the matrix whose columns are the $\mathbf{w}_{k}$, $I$ the infinite
identity matrix and $S$ the infinite (shift) matrix with $1$s on its
superdiagonal, we then have 
\begin{equation*}
PW=W(I+S)
\end{equation*}
and the powers of $P$ satisfy 
\begin{equation*}
P^{n}W=W\sum_{k=0}^{n}\binom{n}{k}S^{k}
\end{equation*}
Note that since $P(k,k+1)=k\alpha _{2}$ in the critical case, we have that 
\begin{equation*}
W(k,k)=\frac{\alpha _{2}^{-\left( k-1\right) }}{(k-1)!},
\end{equation*}
and because $W(k,k)\neq \lambda ^{k}$ for some $\lambda ,$ $W$ cannot be the
transfer matrix of some power series. By construction, the first row of $W$
is $(1,0,\dots )$.

Define $\sigma (k,k^{\prime })=0$ if $k^{\prime }\leq k$ together with 
\begin{equation*}
\sigma (k,k^{\prime })=\sum_{1=i_{1}<i_{2}<\dots <i_{k+1}=k^{\prime
}}\prod_{l=1}^{k}P(i_{l},i_{l+1})\text{ if }k^{\prime }>k\geq 1
\end{equation*}
and, if $k=0$, $\sigma (k,k^{\prime })=\delta _{1,k^{\prime }}$. For
example, $\sigma (2,4)=P(1,2)P(2,4)+P(1,3)P(3,4)$, $\sigma
(2,3)=P(1,2)P(2,3) $. Then, we conjecture that 
\begin{equation*}
W^{-1}(k,k^{\prime })=\sigma (k-1,k^{\prime })\text{, }k^{\prime }\geq k\geq
1.
\end{equation*}
We get:\newline

\textbf{Proposition:} Since we have $x_{n}=\mathbf{e}_{1}^{\prime }P^{n}%
\mathbf{x}$, observing $\mathbf{e}_{1}^{\prime }W=\mathbf{e}_{1}^{\prime
}W^{-1}=\mathbf{e}_{1}^{\prime }$ and $\mathbf{e}_{1}^{\prime }S^{k}=\mathbf{%
e}_{k+1}^{\prime },$%
\begin{equation}
x_{n}=\left( \sum_{k=0}^{n}\binom{n}{k}\mathbf{e}_{k+1}^{\prime }\right)
W^{-1}\mathbf{x}=x+\sum_{k=1}^{n}\binom{n}{k}%
\sum_{k^{\prime }\geq k+1}\sigma (k,k^{\prime })x^{k^{\prime }}.  \label{e7b}
\end{equation}
This constitutes the infinite-dimensional matrix representation of $x_{n}$
in the critical case.\newline

\emph{Slow extinction in the critical case.} With $\phi \left( 0\right) =0$
and $\phi ^{\prime }\left( 0\right) =\alpha _{1}=1$, consider the critical
dynamics, 
\begin{equation*}
x_{n+1}\left( x\right) =\phi \left( x_{n}\left( x\right) \right) \text{, }%
x_{0}\left( x\right) =x.
\end{equation*}
Suppose $x_{n}\left( x\right) $ approaches $0$ as $n$ gets large. An
order-two Taylor development of $\phi $ near $x=0$ therefore gives 
\begin{eqnarray*}
x_{n+1}\left( x\right) &\approx &\phi ^{\prime }\left( 0\right) x_{n}\left(
x\right) +\frac{1}{2}\phi ^{\prime \prime }\left( 0\right) x_{n}\left(
x\right) ^{2} \\
&=&x_{n}\left( x\right) +\frac{1}{2}\phi ^{\prime \prime }\left( 0\right)
x_{n}\left( x\right) ^{2},
\end{eqnarray*}
leading (assuming $\alpha _{2}:=\phi ^{\prime \prime }\left( 0\right) <0$)
to 
\begin{equation}
x_{n}\left( x\right) \sim x/\left( 1-\left( n\alpha _{2}\right) /2\right) ,%
\text{ as }n\text{ is large,}  \label{e7c}
\end{equation}
indeed going to $0$. We conclude:\newline

\textbf{Proposition:} For a critical dynamics (\ref{e1}) for which $\alpha
_{1}=1$, if $\alpha _{2}:=\phi ^{\prime \prime }\left( 0\right) <0$, $%
nx_{n}\left( x\right) \rightarrow 2x/\left( -\alpha _{2}\right) $, and $%
x_{n} $ vanishes at slow algebraic rate $n^{-1}$.

\subsection{Special models arising in population dynamics}

Because we are interested in population dynamics systems for which $%
x_{n}\geq 0$ is the size of some population at time $n$, we shall limit
ourselves to dynamical systems of type (\ref{e1}) generated by $\phi $ with $%
\phi \left( 0\right) =0$ and $\alpha _{1}=\phi ^{\prime }\left( 0\right) >0$%
, $\alpha _{1}\neq 1$ (recall however that the latter construction leading
to (\ref{e5}) and (\ref{e7}) holds even if $\alpha _{1}<0$). The initial
condition $x$ will be assumed to belong to the domain $\left[ 0,x_{b}\right] 
$ where $x_{b}=\inf \left( x>0:\phi \left( x\right) =0\right) \leq x_{c}^{+}$%
, possibly with $x_{b}=\infty $ if $x_{c}^{+}=\infty $. We also need to
assume that the maximal value $\phi ^{*}$ that the function $\phi $ can take
on $\left[ 0,x_{b}\right] $ is $\leq x_{b}$ so that $\phi $ maps $I=\left[
0,x_{b}\right] $ onto $J\subseteq \left[ 0,x_{b}\right] $. If these
conditions hold, the dynamics (\ref{e1}) will be said a population dynamics
model.\newline

\emph{Examples of population models:}

$\left( i\right) $ (logistic map) $\phi \left( x\right) =rx\left( b-x\right) 
$ with $4\geq \alpha _{1}=br>0$, $\alpha _{2}=-r<0$, $x_{b}=b>0$, $%
x_{c}^{+}=\infty .$ Here $I=\left[ 0,b\right] $ and $J=\left[
0,rb^{2}/4\right] .$

$\left( ii\right) $ (homographic map) $\phi \left( x\right) =rx/\left(
1+ax\right) $ with $\alpha _{1}=r>0$, $\alpha _{2}=-ar<0$, $%
x_{b}=x_{c}^{+}=\infty .$ Here $I=\left[ 0,\infty \right] $ and $J=\left[
0,r/a\right] .$

$\left( iii\right) $ (Ricker map) $\phi \left( x\right) =rx\exp \left(
-ax\right) $ with $\alpha _{1}=r>0$, $\alpha _{2}=-ar<0$, $x_{b}=$ $%
x_{c}^{+}=\infty .$ Here $I=\left[ 0,\infty \right] $ and $J=\left[ 0,\phi
^{*}=1/a\right] \subset I.$

$\left( iv\right) $ $\phi \left( x\right) =x_{*}-\sqrt{a\left(
x_{*}-x\right) ^{2}+b}$ with $a,b,x_{*}>0$, $ax_{*}^{2}+b=x_{*}^{2}$. Here $%
I=\left[ 0,2\sqrt{b/\left( 1-a\right) }\right] $ and $J=\left[ 0,\sqrt{%
b/\left( 1-a\right) }\left( 1-\sqrt{1-a}\right) \right] .$ Note $\phi
^{\prime }\left( 0\right) =\alpha _{1}=a<1.$ For this map, $\phi \left(
x\right) =q^{-1}\left( aq\left( x\right) +b\right) $ where $q\left( x\right)
=\left( x-x_{*}\right) ^{2}.$

Note that models $\left( i\right) $ and $\left( iii\right) $ are
single-humped maps, meaning a typically smooth and non-negative function
with exactly one critical point (where $\phi ^{\prime }$ vanishes), a
maximum, and at most one point of inflection to the right of the maximum.

\subsection{Invariant measure of specific maps}

Consider a map $\phi $ from some interval $I$ onto $J\subseteq I$. Then the
invariant measure $\mu $ solves the Perron-Frobenius equation $\mu =\phi
^{-1}\circ \mu ,$ \cite{Ding} and \cite{Las}. When $\phi $ is chaotic, a
density solution $f$ possibly exists, thereby solving $f\left( x\right)
=\int_{I}\delta \left( x-\phi \left( y\right) \right) f\left( y\right) dy$, 
\cite{Thun}. This is also 
\begin{equation}
f\left( x\right) =\sum_{y:\phi \left( y\right) =x}\frac{f\left( y\right) }{%
\left| \phi ^{\prime }\left( y\right) \right| }=\sum_{k=1}^{K\left( x\right)
}f\left( \phi _{k}^{-1}\left( x\right) \right) \left| \phi _{k}^{-1}\left(
x\right) ^{\prime }\right| =\sum_{k=1}^{K\left( x\right) }\frac{f\left( \phi
_{k}^{-1}\left( x\right) \right) }{\left| \phi ^{\prime }\left( \phi
_{k}^{-1}\left( x\right) \right) \right| },  \label{e7d}
\end{equation}
where $K\left( x\right) $ is the number of $\phi -$antecedents of $x$
belonging to the set $\left\{ x:f\left( x\right) >0\right\} $.

Consider now a non-monotone map $\phi $ from some interval $I$ onto itself
(a surjection whose domain and range coincide with $I$). For almost all
initial condition $x$ (drawn at random uniformly on $\left[ 0,1\right] $),
for almost all $\phi ^{\prime }\left( 0\right) $ in a chaotic range of $\phi 
$, a density solution $f$ of (\ref{e7d}) can exist and be made explicit. For
instance, consider the case investigated by \cite{Luev} where in addition: $%
\forall x\in I$, $\phi ^{-1}\left( x\right) $\ is made of $K$\ antecedents
(independently of $x$).

Given $\phi $ and $\kappa \in \Bbb{R}$, consider then the Schr\"{o}der
functional equation, with unknown function $s$, 
\begin{equation*}
s\left( \phi \left( x\right) \right) =\kappa s\left( x\right) .
\end{equation*}
Taking the derivative and then the modulus, $\left| \phi ^{\prime }\left(
x\right) \right| \left| s^{\prime }\left( \phi \left( x\right) \right)
\right| =\left| \kappa \right| \left| s^{\prime }\left( x\right) \right| $.
Thus for each branch $k$, $\left| \phi ^{\prime }\left( \phi _{k}^{-1}\left(
x\right) \right) \right| \left| s^{\prime }\left( x\right) \right| =\left|
\kappa \right| \left| s^{\prime }\left( \phi _{k}^{-1}\left( x\right)
\right) \right| $, showing from (\ref{e7d}), see \cite{Luev}, that if $%
\left| \kappa \right| =K$%
\begin{equation*}
f\left( x\right) =\left| s^{\prime }\left( x\right) \right| .
\end{equation*}
This shows that for chaotic maps $\phi $ mapping some interval $I$ into
itself and for which $\forall x\in I$, $\phi ^{-1}\left( x\right) $\ is made
of $K$\ antecedents (independent of $x$), if it has an invariant density, it
can be explicitly obtained from the solution of a Schr\"{o}der functional
equation. But $s$ can also be related to $h$, the Carleman function of $\phi 
$.\newline

\textbf{Corollary:} Let $\phi $ map $I$ onto itself with $K\left( x\right)
=K $ and $\lambda =\phi ^{\prime }\left( 0\right) >0$. Recalling the
representation (\ref{e7}) of $\phi $: $h\left( \phi \left( x\right) \right)
=\lambda h\left( x\right) $, with $\alpha =\log _{K}\lambda $, we get $%
s\left( x\right) =h\left( x\right) ^{1/\alpha }$ and so $f\left( x\right) =%
\frac{1}{\left| \alpha \right| }\left| h\left( x\right) ^{1/\alpha
-1}h^{\prime }\left( x\right) \right| .$\newline

This shows that in some cases, (\ref{e7}) can also be useful for the
computation of the invariant density of $\phi $. Let us illustrate these
facts for the logistic map:

- With $I=\left[ 0,1\right] $, consider the logistic map $\phi \left(
x\right) =rx\left( 1-x\right) $ with $b=1$ and $r=2$. For this map, $K=2$
but $\phi $ maps $\left[ 0,1\right] $ to $\left[ 0,1/2\right] \subset \left[
0,1\right] $. Then, for every $x\in \left( 0,1/2\right) $, $\phi \left(
x\right) =h^{-1}\left( 2h\left( x\right) \right) $ where $h\left( x\right)
=-\log \left( 1-2x\right) /2$ and $h^{-1}\left( x\right) =\left(
1-e^{-2x}\right) /2$, the inverse of $h$. The invariant measure is a Dirac
at the fixed point $x=1/2$ of $\phi $. If $2<r<r\simeq 3.5699456$ (the
Feigenbaum constant), the invariant measure is a Dirac measure concentrating
at the fixed points obtained in the period doubling process of the logistic
map.\newline

- Consider the logistic map $\phi \left( x\right) =rx\left( 1-x\right) $
with $b=1$ and $r=4$: $\phi $ maps $\left[ 0,1\right] $ into $\left[
0,1\right] $ and $K=2$ for all $x$. Then $\phi \left( x\right) =h^{-1}\left(
4h\left( x\right) \right) $ where $h\left( x\right) =\arcsin \left( \sqrt{x}%
\right) ^{2}$ and $h^{-1}\left( x\right) =\sin \left( \sqrt{x}\right) ^{2}$%
\footnote{%
In this case, the inverse function of $h\left( x\right) =\arcsin \left( 
\sqrt{x}\right) ^{2}$ is $\sin \left( \sqrt{x}\right) ^{2}$, restricted to
the interval $\left[ 0,\pi /2\right] $ whereas its analytic continuation $%
h^{-1}$ is the same function now defined on the whole positive real line
(its maximal domain of convergence).}. The invariant measure is $\mu \left(
dx\right) =f\left( x\right) dx$ where $f\left( x\right) =\pi ^{-1}\left(
x\left( 1-x\right) \right) ^{-1/2}=\left| h^{1/2}\left( x\right) ^{\prime
}\right| =\left| s^{\prime }\left( x\right) \right| $ where $s\left(
x\right) =\arcsin \left( \sqrt{x}\right) $ solves $s\left( \phi \left(
x\right) \right) =2s\left( x\right) .$

We now wish to further investigate some aspects of the invariant density
problem for the general logistic map.

\section{Invariant densities for $\phi(x) = r x (1-x)$}

Let us further consider the logistic map $\phi \left( x\right) =rx\left(
1-x\right) $ with $b=1$ and assume now $4>r=\lambda >2$. This map is not
into $I$ but onto only. In this case the dynamics moves to the interval $%
\left[ \phi \left( r/4\right) ,r/4\right] $ in finite time and stays there%
\footnote{$x>r/4$ has no antecedent and cannot be reached, the function is
strictly increasing on $\left] 0, \phi(r/4)\right]$, and the interval $%
\left[\phi(r/4),r/4 \right]$ is mapped onto itself.}: the support of the
invariant measure is expected to lie within this interval and, whenever $%
4\geq r>3.5699456$ (the Feigenbaum constant for the onset of chaos), the set
of points with positive measure possibly consists in a finite union of
disconnected subintervals embedded within this support. Within the interval $%
r\in \left[ 3.5699456...,4\right] $, it is widely known that there are
subintervals where the map $\phi $ is not chaotic (has negative Lyapounov
exponent) corresponding to $x_{n}$ reaching a cycle of any length. In these
subintervals, the invariant measure is expected to be a Dirac measure.%
\newline

The purpose of this section is to exhibit some features of the invariant
density in the chaotic regimes, when this density exists. In the following,
we will use the notation $\alpha_n(r) := \phi_n(1/2)$.

To illustrate this section, we will often refer to a specific value of $r$, $%
r \simeq 3.6785735$, which is the one studied by Ruelle in \cite{Ruelle}.
This value is the smallest $r$ for which there is a density with full
support $\left[ 1/r,r/4\right] $, and it solves $\left( r-2\right) \left(
r^{2}-4\right) =16$.

\subsection{Divergence points of the invariant density}

When $r=4$, the invariant density diverges likes $x^{-1/2}$ near $0$ and $%
\left( 1-x\right) ^{-1/2}$ near $1$, $\left\{ 0,1\right\} $ being the
boundary of the support of the invariant density. When $r=4$, $\phi $ maps $%
1/2$ to $1$ and then to $0$, a fixed point of the dynamics: $\alpha
_{0}\left( r\right) =1/2$, $\alpha _{1}\left( r\right) =1$ and $\alpha
_{2}\left( r\right) =0$. The density diverges at $\left\{ 0,1\right\} $, the
successive image of $1/2$ under $\phi $.

When $r\simeq 3.6785735$, $\phi $ maps $1/2$ to $r/4$ and then to $1/r$ and
finally to $1-1/r$, a fixed point of the dynamics: $\alpha _{0}\left(
r\right) =1/2$, $\alpha _{1}\left( r\right) =r/4$, $\alpha _{2}\left(
r\right) =1/r$ and $\alpha _{3}\left( r\right) =1-1/r$. The density is
expected to diverge at points $\left\{ 1/r,1-1/r,r/4\right\} $, the
successive images of $1/2$ under $\phi $.

This seems to be a generic fact.

For a generic value of $r$, when $x=\alpha _{1}\left( r\right) =r/4$, there
is only one antecedent, namely $y=1/2$\ and there, $\phi ^{\prime }\left(
y=1/2\right) =0$, so, in view of (\ref{e7d}), the invariant density $f\left(
x\right) $ of the mass at $x=r/4$\ must diverge, if it exists. This is also
true for $x=\phi _{1}\left( r/4\right) =\phi _{2}\left( 1/2\right) =\alpha
_{2}\left( r\right) $: for this $x$ indeed, $y=r/4$\ is an antecedent and
because $f\left( y\right) $\ diverges for this value of $y$, so does $%
f\left( x\right) $.\emph{\ }By induction,\newline

\textbf{Proposition:} For all $x=\alpha _{n}\left( r\right) $, $n\geq 1$,
there is a divergence of the invariant density.\newline

For the Ruelle value of $r$, the sequence $\phi _{n}\left( 1/2\right)
=\alpha _{n}\left( r\right) $, $n\geq 1$, stops (converges) after three
steps, so one expects a divergence of the density $f$ in that point which is 
$1-1/r$\ (the fixed point of $\phi $) but also at the intermediate points $%
\alpha _{1}\left( r\right) =r/4$, $\alpha _{2}\left( r\right) =1/r$.
Following this argument, when $r\simeq 3.89087$ (defined by $\alpha_5(r) =
\alpha_3(r) \neq \alpha_4(r)$), one expects a divergence of the density at
the four points $\alpha _{1}\left( r\right) =r/4=\phi _{1}\left( 1/2\right) $%
, $\alpha _{2}\left( r\right) =\phi _{2}\left( 1/2\right) $, $\alpha
_{3}\left( r\right) =\phi _{3}\left( 1/2\right) $ and $\alpha _{4}\left(
r\right) =\phi _{4}\left( 1/2\right) $. The sequence stops there because for
this value of $r$, $\alpha _{5}\left( r\right) =\alpha _{3}\left( r\right) $%
, $\alpha _{6}\left( r\right) =\alpha _{4}\left( r\right) $ and no new $%
\alpha _{n}\left( r\right) $ is expected, this sequence oscillating between $%
\alpha _{3}\left( r\right) $ and $\alpha _{4}\left( r\right) $ with $\alpha
_{3}\left( r\right) >\alpha _{4}\left( r\right) $. This concerns special
values of $r$ for which $f$ has full support and finitely many peaks.\newline

This poses the following general question: let $\mathcal{C}_{m}$ be a cycle
of length $m$ of $\phi $. Consider the set 
\begin{equation*}
\mathcal{R}:=\left\{ r\text{ in the chaotic regime: }\exists n,m<\infty 
\text{: }\alpha _{n}\left( r\right) \in \mathcal{C}_{m}\text{ and }\alpha
_{n-1}\left( r\right) \notin \mathcal{C}_{m}\right\} .
\end{equation*}
If $r\in \mathcal{R}$ and if corresponding map $\phi $ admits a density $f$,
then there must be $n+m-1$ peaks where $f$ diverges, by the arguments above.
If $r\notin \mathcal{R}$ and if $f$ exists, then $f$ has infinitely many
peaks within its support. We don't know if $\mathcal{R}$ is dense within the
subset of values of $r$ in the chaotic regime, nor do we have a specific
value of $r\notin \mathcal{R}$. For an $r\notin \mathcal{R}$, the sequence $%
\left\{ \alpha _{n}\left( r\right) \right\} $ does not enter a cycle in
finite time and one expects a density diverging on this countable set values
for $x$ within $\left[ \phi \left( r/4\right) ,r/4\right] $.\newline

It remains to determine the type of the divergence of $f$. We shall develop
our arguments for the Ruelle value of $r$. We have 
\begin{equation*}
\phi \left( y\right) =\phi \left( \frac{1}{2}+\left( y-\frac{1}{2}\right)
\right) =\frac{r}{4}-r\left( y-\frac{1}{2}\right) ^{2}\text{, }\phi ^{\prime
}\left( y\right) =r\left( 1-2y\right) .
\end{equation*}
Put $x=r/4-\varepsilon $ and $y=1/2-\varepsilon ^{\prime }$ with $\epsilon,
\epsilon^{\prime} >0$. Because $\phi ^{\prime }\left( 1/2\right) =0$, we
have $\left| \phi ^{\prime }\left( y\right) \right| =2r\varepsilon ^{\prime
} $ and $\phi \left( y\right) =r/4-r\varepsilon ^{\prime
2}=x=r/4-\varepsilon $. Thus $r\varepsilon ^{\prime 2}=\varepsilon $.

In view of 
\begin{equation*}
f\left( x\right) =\sum_{y:\phi \left( y\right) =x}\frac{f\left( y\right) }{%
\left| \phi ^{\prime }\left( y\right) \right| },\text{ therefore }f\left(
r/4-\varepsilon \right) \sim \frac{f\left( 1/2\right) }{2\sqrt{r\varepsilon }%
}.
\end{equation*}
Similarly, with some linear relation between $\epsilon $, $\epsilon ^{\prime
}$ and $\epsilon ^{\prime \prime }$, the antecedents of $1/r+\epsilon $ are $%
r/4-\epsilon ^{\prime }$ and $1-r/4+\epsilon ^{\prime \prime }$, the latter
being outside of the support of $f$ so its value there is zero. The
antecedents of $1-1/r+\epsilon $ are $1/r+\epsilon ^{\prime }$ and $%
1-1/r-\epsilon ^{\prime \prime }$. The antecedents of $1-1/r-\epsilon $ are $%
1-1/r+\epsilon ^{\prime }$ and $1/r-\epsilon ^{\prime \prime }$, the latter
being outside of the support of $f$. With these relations, we find that 
\begin{equation*}
(1/r+\epsilon )\sim a/\sqrt{\epsilon },\text{ }f(1-1/r-\epsilon )\sim b/%
\sqrt{\epsilon },\text{ }f(1-1/r+\epsilon )\sim c/\sqrt{\epsilon }
\end{equation*}
with 
\begin{equation*}
a:=\frac{f(1/2)}{\left| \phi ^{\prime }(1/r)\right| \left| \phi ^{\prime
}(1/r)\right| ^{1/2}},\text{ }b:=\frac{a\left| \phi ^{\prime }(1/r)\right| }{%
\left| \phi ^{\prime }(1/r)\right| -1},\text{ }c:=\frac{a\left| \phi
^{\prime }(1/r)\right| ^{3/2}}{\left| \phi ^{\prime }(1/r)\right| -1}.
\end{equation*}
\textbf{Proposition:} There is an algebraic divergence of $f$ at the peaks
of order $-1/2$ whereby $f$ is integrable.

\subsection{Disconnected invariant measure support}

As a polynomial in the variable $r$, $\alpha _{n}\left( r\right)$ has degree 
$2^{n}-1$. When $r$ is larger than the Ruelle value $r\simeq 3.6785735$,
whenever the regime is chaotic, the support of the invariant measure is made
of one single piece, namely the full interval $\left[ \alpha _{2}\left(
r\right) =\phi \left( r/4\right) ,\alpha _{1}\left( r\right) =r/4\right] $.
The above value of $r$ is when $\alpha _{3}\left( r\right) $ is a fixed
point of $\phi $, namely $\alpha _{3}\left( r\right) =\phi \left( \alpha
_{3}\left( r\right) \right) =\alpha _{4}\left( r\right) $. Because $\alpha
_{3}\left( r\right) =1-1/r$ and $\phi \left( 1/r\right) =1-1/r,$ this Ruelle
value of $r$ is also obtained when $\alpha _{2}\left( r\right) =r^{2}\left(
1-r/4\right) /4=1/r$, indeed leading to $r\simeq 3.6785735.$ When $r$
becomes slightly less than the Ruelle value, the support of the invariant
density of $\phi $ splits in the two pieces $\left[ \alpha _{2}\left(
r\right) ,\alpha _{4}\left( r\right) \right] \cup \left[ \alpha _{3}\left(
r\right) ,\alpha _{1}\left( r\right) \right] $, because $\alpha _{4}\left(
r\right) <\alpha _{3}\left( r\right) $. These two pieces each split again in
two additional pieces for the value of $r$ for which $\alpha _{5}\left(
r\right) =\alpha _{7}\left( r\right) $ and $\alpha _{6}\left( r\right)
=\alpha _{8}\left( r\right) $, corresponding to $\alpha _{5}\left( r\right) $
and $\alpha _{6}\left( r\right) $ being respectively one of the two known
fixed points of $\phi _{2} $ (which are not the fixed points $\left\{
0,1-1/r\right\} $ of $\phi _{1}=\phi $). Slightly below this value of $r$,
the support of the invariant density of $\phi $ splits into the four pieces $%
\left[ \alpha _{2}\left( r\right) ,\alpha _{8}\left( r\right) \right] \cup
\left[ \alpha _{6}\left( r\right) ,\alpha _{4}\left( r\right) \right] \cup
\left[ \alpha _{3}\left( r\right) ,\alpha _{7}\left( r\right) \right] \cup
\left[ \alpha _{5}\left( r\right) ,\alpha _{1}\left( r\right) \right] $:
this transition is seen to occur at $r\simeq 3.5925722$. The next third step
generates the $2^{3}=8$ pieces 
\begin{eqnarray*}
&&\left[ \alpha _{2}\left( r\right) ,\alpha _{16}\left( r\right) \right]
\cup \left[ \alpha _{12}\left( r\right) ,\alpha _{8}\left( r\right) \right]
\cup \left[ \alpha _{6}\left( r\right) ,\alpha _{14}\left( r\right) \right]
\cup \left[ \alpha _{10}\left( r\right) ,\alpha _{4}\left( r\right) \right]
\cup \\
&&\left[ \alpha _{3}\left( r\right) ,\alpha _{15}\left( r\right) \right]
\cup \left[ \alpha _{11}\left( r\right) ,\alpha _{7}\left( r\right) \right]
\cup \left[ \alpha _{5}\left( r\right) ,\alpha _{13}\left( r\right) \right]
\cup \left[ \alpha _{9}\left( r\right) ,\alpha _{1}\left( r\right) \right]
\end{eqnarray*}
whenever $\alpha _{16}\left( r\right) =\alpha _{12}\left( r\right) $, $%
\alpha _{14}\left( r\right) =\alpha _{10}\left( r\right) $, $\alpha
_{15}\left( r\right) =\alpha _{11}\left( r\right) $ and $\alpha _{13}\left(
r\right) =\alpha _{9}\left( r\right) $ corresponding to $\alpha _{12}\left(
r\right) ,$ $\alpha _{10}\left( r\right) $, $\alpha _{11}\left( r\right) $
and $\alpha _{9}\left( r\right) $ being the four fixed points of $\phi _{4}$
which are neither the ones of $\phi _{1}$ nor the ones of $\phi _{2}$. For
some special value of $r$ which can be computed, the four pieces split into
these eight pieces.

This binary splitting process of the support can be iterated until one
reaches the critical Feigenbaum value of $r$, namely $r\simeq 3.5699456$
where the invariant measure is expected to be singular (whose support
consists in an uncountable number of points). For values of $r$ less than
this Feigenbaum value and larger than three, the invariant measure is a
Dirac measure concentrated on the period two cyclic points appearing in the
period doubling process onsetting after $r=3$.\newline

\subsection{Invariant measure for some specific values of $r$}

Let us consider the Ruelle value of $r$. For this value of $r$, the quartic
map 
\begin{equation*}
\psi \left( x\right) =\left( r-2\right) ^{2}x\left( 1-x\right) \left[
1+\left( r-2\right) x\left( 1-x\right) \right]
\end{equation*}
maps $\left[ 0,1\right] $ into $\left[ 0,1\right] $ and it has $K\left(
x\right) =2$ branches for each $x\in \left[ 0,1\right] \setminus \{1/2\} .$
Furthermore, $\psi \left( x\right) =q_{2}\left( x\right) :=q\circ q\left(
x\right) $ where $q\left( x\right) =\left( r-2\right) x\left( x-1\right) $
and, with $a\left( x\right) =\left( r-1-rx\right) /\left( r-2\right) $%
\begin{equation*}
\phi \left( x\right) =a^{-1}\left( q\left( a\left( x\right) \right) \right) 
\text{ and }\phi _{2}\left( x\right) :=\phi \circ \phi \left( x\right)
=a^{-1}\left( \psi \left( a\left( x\right) \right) \right) .
\end{equation*}
As stated before, this value is also one for which $\alpha _{3}\left(
r\right) =\alpha _{4}\left( r\right) $ ($\alpha _{3}\left( r\right) $ is a
fixed point of $\phi $).

Let now $h$ be the Carleman function associated to $\psi $, defined by $\psi
\left( x\right) =h^{-1}\left( \lambda h\left( x\right) \right) $, $\lambda
=\psi ^{\prime }\left( 0\right) =\left( r-2\right) ^{2}>2$, as from (\ref{e7}%
). The dynamical system generated by $\psi $ has an absolutely continuous
invariant measure, given by $f_{\psi }\left( x\right) =\left| s^{\prime
}\left( x\right) \right| $ where $s\left( \psi \left( x\right) \right)
=2s\left( x\right) $ and $s\left( x\right) =h^{1/\alpha }\left( x\right) $, $%
\alpha =\log _{2}\lambda >1.$ Therefore so does the dynamical system
generated by $\phi _{2}$ with invariant density given by $f_{\phi
_{2}}\left( x\right) =\left| a^{\prime }\left( x\right) \right| f_{\psi
}\left( a\left( x\right) \right) $, with support $\left[ 1/r,1-1/r\right] .$
And then so does the dynamical system generated by $\phi $ itself, with the
density $f_{\phi }\left( x\right) =\frac{1}{2}\left( f_{\phi _{2}}\left(
x\right) +\phi ^{-1}\circ f_{\phi _{2}}\left( x\right) \right) $, where $%
\phi ^{-1}\circ f_{\phi _{2}}\left( x\right) =\left| \phi ^{\prime }\left(
x\right) \right| f_{\phi _{2}}\left( \phi \left( x\right) \right) $ is the
image density of $f_{\phi _{2}}$ under $\phi ^{-1}$. The support of $f_{\phi
}$ is $\left[ 1/r,r/4\right] =\left[ 1/r,1-1/r\right] \cup \phi ^{-1}\left[
1/r,1-1/r\right] $. We conclude that there is a Carleman inspired analytic
expression of the invariant density for the logistic model when $r$ takes on
the Ruelle value. This is not the only one case.\newline

Looking for an affine function $a\left( x\right) =c-dx$ such that $\phi
_{4}\left( x\right) =a^{-1}\left( \Psi \left( a\left( x\right) \right)
\right) $ for some map $\Psi =\psi \circ \psi $ with $\psi \left( x\right)
=ax\left( 1-x\right) \left[ 1+bx\left( 1-x\right) \right] $, such that $\Psi 
$ maps $\left[ 0,1\right] $ onto itself, leads to another solution to the
above Ruelle scenario, namely: $a\simeq -1.7214305$, $b\simeq 0.1592945$, $%
c\simeq -2.2648957$, $d\simeq -5.5297915$ and $r\simeq 3.5925722$. The
latter is the value of $r$ where the support of the invariant measure splits
from 2 pieces to 4. In that case, we have $\lambda :=\Psi ^{\prime
}(0)=a^{2} $, $\Psi $ having two branches on $\left[ 0,1\right] \setminus
\{1/2\}$. Using the same Carleman matrix approach as previously stated, it
is possible to compute the invariant measure associated to the mapping $\Psi 
$ (now with $h$ the Carleman function of $\Psi $ and $s\left( x\right)
=h^{1/\alpha }\left( x\right) $ for $\alpha =\log _{2}a^{2}$). From that
result and following the arguments developed for the Ruelle value, one can
clearly compute the invariant density first of the function $\phi _{4}$ on a
given subset of its support, namely $f_{\phi _{4}}$, and then get the
invariant density of $\phi $ on its support by 
\begin{equation*}
f_{\phi }=\frac{1}{4}\left( f_{\phi _{4}}+\phi ^{-1}\circ f_{\phi _{4}}+\phi
^{-2}\circ f_{\phi _{4}}+\phi ^{-3}\circ f_{\phi _{4}}\right) .
\end{equation*}
\textbf{Proposition:} For the particular value of $r\simeq 3.5925722$, the
invariant density of $\phi $ can be computed.\newline

Although for $r\simeq 3.6785735$ and for $r\simeq 3.5925722$, the maps $\phi 
$ were not from $\left[ 0,1\right] $ onto $\left[ 0,1\right] $, the powers $%
\phi _{2}$ and $\phi _{4}$ of these maps were shown to be affine-conjugate
to maps $\psi $ and $\Psi $ which now map $\left[ 0,1\right] $ onto $\left[
0,1\right] $, thereby amenable to the Schr\"{o}der and Carleman
representation of the invariant densities.\newline

\section{The general case $\alpha _{0}\neq 0$}

Let $\phi _{0}\left( x\right) =\sum_{k\geq 0}\alpha _{k}x^{k}=\alpha
_{0}+\phi \left( x\right) $, now with $\alpha _{0}=\phi _{0}\left( 0\right)
\neq 0.$ Consider now the dynamical system 
\begin{equation}
x_{n+1}=\phi _{0}\left( x_{n}\right) \text{, }x_{0}=x.  \label{e8}
\end{equation}
Let

\begin{equation}
P_{0}\left( k,k^{\prime }\right) =\left[ x^{k^{\prime }}\right] \phi
_{0}\left( x\right) ^{k}\text{, }k,k^{\prime }\geq 0.  \label{e9}
\end{equation}
Note $P_{0}$ no longer is triangular. Specifically, for $k,k^{\prime }\geq 1$%
\begin{eqnarray*}
P_{0}\left( k,k^{\prime }\right) &=&\left[ x^{k^{\prime }}\right] \left(
\alpha _{0}+\phi \left( x\right) \right) ^{k}=\sum_{l=0}^{k}\binom{k}{l}%
\alpha _{0}^{k-l}\left[ x^{k^{\prime }}\right] \phi \left( x\right) ^{l} \\
&=&\sum_{l=1}^{k}\binom{k}{l}\alpha _{0}^{k-l}P\left( l,k^{\prime }\right)
=\sum_{l=1}^{k}\binom{k}{l}\alpha _{0}^{k-l}\widehat{B}_{k^{\prime
},l}\left( \alpha _{1},\alpha _{2},...,\alpha _{k^{\prime }-l+1}\right) ,
\end{eqnarray*}
together with $P_{0}\left( 0,k^{\prime }\right) =\delta _{0,k^{\prime }}$, $%
P_{0}\left( k,0\right) =\alpha _{0}^{k}$, for $k\geq 0.$

Hence $P_{0}=\overline{B}\overline{P}$ where $\overline{B}\left( k,l\right) =%
\binom{k}{l}\alpha _{0}^{k-l}$, $k\geq l\geq 0$ and $\overline{P}=\left[ 
\begin{array}{cccc}
1 & 0 & \cdots & \cdots \\ 
0 &  &  &  \\ 
\vdots &  & P &  \\ 
\vdots &  &  & 
\end{array}
\right] ,$ so with $P_{0}=\left[ 
\begin{array}{cccc}
1 & 0 & \cdots & \cdots \\ 
\alpha _{0} &  &  &  \\ 
\alpha _{0}^{2} &  & BP &  \\ 
\vdots &  &  & 
\end{array}
\right] $ and $B\left( k,l\right) =\binom{k}{l}\alpha _{0}^{k-l}$, $k\geq
l\geq 1$, the incomplete lower-triangular Pascal (binomial) matrix. Note $%
\left( \overline{B}\mathbf{x}\right) _{k}=\left( \alpha _{0}+x\right) ^{k}$:
the full Pascal matrix $\overline{B}$ is the Carleman matrix of the shift
function $b\left( x\right) =\alpha _{0}+x$. It also holds $\overline{B}%
\left( k,k\right) =1$, $k\geq 0$ and $P_{0}$ admits a LU factorization.

Then, with $\mathbf{e}_{1}^{\prime }=\left( 0,1,0,0,...\right) $ and $%
\overline{\mathbf{x}}^{\prime }=\left( 1,x,x^{2},...\right) $, 
\begin{equation}
x_{n}=\mathbf{e}_{1}^{\prime }P_{0}^{n}\overline{\mathbf{x}}\mathbf{.}
\label{e10}
\end{equation}
This shows that a general non-linear dynamical system (\ref{e8}) generated
by $\phi _{0}$ with $\phi _{0}\left( 0\right) \neq 0$ is also a linear
infinite-dimensional system with complete `transfer matrix' $P_{0}$. Because 
$P_{0}$ no longer is triangular, its eigenvalues are not known nor is its
diagonalization easy, if even possible. So $P_{0}^{n}$ is very complex. Note
in particular that $x_{n}\left( x=0\right) =\mathbf{e}_{1}^{\prime }P_{0}^{n}%
\mathbf{e}_{0}=P_{0}^{n}\left( 1,0\right) $, the $\left( 1,0\right) -$entry
of $P_{0}^{n}\mathbf{.}$

In some cases however, the general problem (\ref{e8}) can be taken back to a
simpler problem of type (\ref{e1}), see \cite{GK}.

Let $y_{n}=x_{n}-\rho $ for some real number $\rho $. Then 
\begin{equation*}
y_{n+1}=\phi _{0}\left( y_{n}+\rho \right) -\rho \text{, }y_{0}=x-\rho .
\end{equation*}
Suppose there is a real number $\rho $ such that $\phi _{0}\left( \rho
\right) =\rho $ (the map $\phi _{0}$ has a fixed point). Then 
\begin{equation}
y_{n+1}=\overline{\phi }\left( y_{n}\right) :=\phi _{0}\left( y_{n}+\rho
\right) -\rho \text{, }y_{0}=x-\rho ,  \label{e11}
\end{equation}
where $\overline{\phi }\left( 0\right) =\phi _{0}\left( \rho \right) -\rho
=0.$ If this is the case, the new system (\ref{e11}) now generated by $%
\overline{\phi }$ is of the form (\ref{e1}). It can be solved as a (\ref{e1}%
) model with the new

\begin{equation*}
\overline{\phi }\left( y\right) =\sum_{k\geq 0}\alpha _{k}\left( y+\rho
\right) ^{k}-\rho =:\sum_{l\geq 1}y^{l}\overline{\alpha }_{l},
\end{equation*}
where $l!\overline{\alpha }_{l}=\sum_{k\geq l}\frac{\rho ^{k-l}}{\left(
k-l\right) !}\left( k!\alpha _{k}\right) $ is of convolution type. And then,
assuming $\overline{\lambda }:=\overline{\alpha }_{1}\neq 1$, 
\begin{equation}
x_{n}=\overline{h}^{-1}\left( \overline{\lambda }^{n}\overline{h}\left(
x-\rho \right) \right) +\rho ,  \label{e12}
\end{equation}
where $\overline{h}$ is associated to the new generator $\overline{\phi }$
(just like $h$ in (\ref{e7}) was to $\phi $ in (\ref{e1})). Introducing $%
\widetilde{h}\left( x\right) =\overline{h}\left( x-\rho \right) $, this is
also $x_{n}=\widetilde{h}^{-1}\left( \overline{\lambda }^{n}\widetilde{h}%
\left( x\right) \right) $ similar to (\ref{e7}), except that $\widetilde{h}%
\left( 0\right) \neq 0$. Note also that $\overline{\phi }^{\prime }\left(
0\right) =:\overline{\alpha }_{1}=\phi _{0}^{\prime }\left( \rho \right) $
is not necessarily $>0.$ And depending on $\left| \phi _{0}^{\prime }\left(
\rho \right) \right| <1$ (or $>1$), $\rho $ is a stable (unstable) fixed
point of (\ref{e8}). Equivalently, (see e.g. \cite{GK})\newline

\textbf{Proposition:} With $Q$ the upper-triangular Carleman matrix of $%
\overline{\phi }$ (easily diagonalizable with $VQ=D_{\overline{\mathbf{%
\lambda }}}V$ where $D_{\overline{\mathbf{\lambda }}}=$diag$\left( 1,%
\overline{\lambda },\overline{\lambda }^{2},...\right) $), the Carleman
matrix $P_{0}$ of $\phi _{0}$ therefore obeys 
\begin{equation}
P_{0}=B_{\rho }^{-1}QB_{\rho }^{{}}=\left( VB_{\rho }\right) ^{-1}D_{%
\overline{\mathbf{\lambda }}}VB_{\rho },  \label{e13}
\end{equation}
where $B_{\rho }$ is the lower-triangular Carleman matrix of the shift
function $b_{\rho }\left( x\right) =x-\rho $: $B_{\rho }\left( k,l\right)
=\left( -1\right) ^{k-l}\binom{k}{l}\rho ^{k-l}$ and $B_{\rho }^{-1}\left(
k,l\right) =\binom{k}{l}\rho ^{k-l}$.

\emph{Remarks:}

- The only Carleman matrices which are lower-triangular are the ones
associated to an affine map as the one above.

- With $\mathbf{\phi }_{0}\left( \rho \right) ^{\prime }=\left( 1,\phi
_{0}\left( \rho \right) ,\phi _{0}\left( \rho \right) ^{2},...\right) $ and $%
\mathbf{\rho }^{\prime }=\left( 1,\rho ,\rho ^{2},...\right) $, for all
fixed point $\rho $ of $\phi _{0}:$ $P_{0}\mathbf{\rho }=\mathbf{\phi }%
_{0}\left( \rho \right) =\mathbf{\rho }$ showing that $\mathbf{\rho }$ is a
right eigenvector of $P_{0}$ associated to its eigenvalue $1$.

- If the map $\phi _{0}$ has more than one real fixed point, the latter
construction holds for any of these fixed points, showing that (\ref{e12})
is not unique.

- If the map $\phi _{0}$ has no real fixed point, $P_{0}$ is not
real-diagonalizable.

For example, the matrix $P_{0}$ associated to the map $\phi _{0}\left(
x\right) =x+\alpha _{0}/\left( x+1\right) $ with $\phi _{0}\left( 0\right)
=\alpha _{0}\neq 0$ is not diagonalizable. However, this model has a fixed
point at infinity. Exchanging $\infty $ and $0$ can be done while using the
transformation $y=1/x.$ The dynamics for the $y$s is thus $y_{n+1}=\phi
\left( y_{n}\right) $, $y_{0}=y=1/x$, with 
\begin{equation*}
\phi \left( y\right) =1/\phi _{0}\left( 1/y\right) =\frac{y\left( 1+y\right) 
}{1+y+\alpha _{0}y^{2}},
\end{equation*}
now with a fixed point at $y=0$ and of type (\ref{e1}), with $\phi $ rational%
$.$ We have $\phi ^{\prime }\left( 0\right) =\alpha _{1}=1$ and $\phi
^{^{\prime \prime }}\left( 0\right) =\alpha _{2}=0,$ a critical model,
therefore amenable to Jordanization.

Another example is $\phi _{0}\left( x\right) =x+x^{2}+\alpha _{0}$ with $%
\phi _{0}\left( 0\right) =\alpha _{0}\neq 0$. This map has no real fixed
point but it has two complex fixed points $\rho =\pm i\sqrt{c}$. The matrix $%
P_{0}$ associated to this map $\phi _{0}$ is not real-diagonalizable but it
is complex-diagonalizable. In such cases, only (\ref{e10}) holds, but not (%
\ref{e12}) where $\overline{h}$ is real-valued.

\subsection{An equivalent conjugation representation of $\phi _{0}$\ having
a fixed point}

Assuming $\overline{\lambda }:=\overline{\alpha }_{1}=\phi _{0}^{\prime
}\left( \rho \right) \neq 1$, with $\widetilde{h}\left( x\right) =\overline{h%
}\left( x-\rho \right) $, we obtained 
\begin{equation*}
x_{1}=\phi _{0}\left( x\right) =\overline{h}^{-1}\left( \overline{\lambda }%
\text{ }\overline{h}\left( x-\rho \right) \right) +\rho =\widetilde{h}%
^{-1}\left( \overline{\lambda }\widetilde{h}\left( x\right) \right) ,
\end{equation*}
where $\overline{h}$ is associated to the generator $\overline{\phi }$.
Define $g\left( x\right) :=\widetilde{h}\left( x\right) -\widetilde{h}\left(
0\right) =\overline{h}\left( x-\rho \right) -\overline{h}\left( -\rho
\right) $, now obeying $g\left( 0\right) =0$ and $g\left( \rho \right) =-%
\overline{h}\left( -\rho \right) .$ Clearly then, with $\overline{\mu }%
:=\left( \overline{\lambda }-1\right) \overline{h}\left( -\rho \right)
=\left( 1-\overline{\lambda }\right) g\left( \rho \right) ,$ 
\begin{equation}
x_{1}=\phi _{0}\left( x\right) =g^{-1}\left( \overline{\lambda }g\left(
x\right) +\overline{\mu }\right) ,  \label{e14}
\end{equation}
showing that $\phi _{0}$ is $g-$conjugate to the affine function $\overline{%
\lambda }x+\overline{\mu }$, with $c=\phi _{0}\left( 0\right) =g^{-1}\left( 
\overline{\mu }\right) >0$. This can be iterated to give 
\begin{equation}
x_{n}=g^{-1}\left( \overline{\lambda }_{n}g\left( x\right) +\overline{\mu }%
_{n}\right) ,\text{ with }\overline{\lambda }_{n}=\overline{\lambda }^{n}%
\text{ and }\overline{\mu }_{n}=\overline{\mu }\left( 1+\overline{\lambda }%
+...+\overline{\lambda }^{n-1}\right) =g\left( \rho \right) \left( 1-%
\overline{\lambda }^{n}\right) .  \label{e15}
\end{equation}
If $\left| \overline{\lambda }\right| <1$ ($\rho $ is a stable fixed point
of $\phi _{0}$), $x_{n}\rightarrow \rho $\emph{.}

\subsection{Population models with immigration}

In population dynamics systems for which $x_{n}\geq 0$, we shall limit
ourselves to dynamical systems of type (\ref{e8}) generated by $\phi _{0}$
with $\phi _{0}\left( 0\right) =c>0$. No need to require here anymore that $%
\phi _{0}^{\prime }\left( 0\right) =\phi ^{\prime }\left( 0\right) =\alpha
_{1}>0$. The initial condition $x$ will be assumed to belong to the domain $%
\left[ 0,x_{b}\right] $ where $x_{b}=\inf \left( x>0:\phi _{0}\left(
x\right) =0\right) $, possibly with $x_{b}=\infty $. Here $\phi _{0}\left(
0\right) >0$ interprets as an immigration rate. We also need to assume that
the maximal value $\phi _{0}^{*}$ that $\phi _{0}\left( x\right) $ can take
on $\left[ 0,x_{b}\right] $ is $\leq x_{b}$ so that $\phi _{0}$ maps $%
I=\left[ 0,x_{b}\right] $ onto $J\subseteq I$. These $\phi _{0}$ are
amenable to the formalism (\ref{e10}) and (\ref{e14}). \newline

\emph{Examples of population models with immigration }$\phi _{0}\left(
0\right) =c>0$\emph{:}

$\left( i\right) $ (logistic map) $\phi _{0}\left( x\right) =c+rx\left(
b-x\right) $, $r>0$, $x_{b}=\left( br+\sqrt{\Delta }\right) /\left(
2r\right) >b>0$, $\Delta =\left( br\right) ^{2}+4rc$, $x_{c}^{+}=\infty .$
Here $I=\left[ 0,x_{b}\right] $ and $J=\left[ 0,c+rb^{2}/4\right] \subseteq
I $ provided $\phi _{0}^{*}=c+rb^{2}/4\leq x_{b}.$

$\left( i^{\prime }\right) $ (logistic map') $\phi _{0}\left( x\right)
=c+rx\left( b-x\right) $, $r<0$, $b>0$. In this setup, $\phi _{0}^{\prime
}\left( 0\right) =\phi ^{\prime }\left( 0\right) =\alpha _{1}=rb<0$

- If $\phi _{0}\left( b/2\right) =c+r\left( b/2\right) ^{2}>0$, then $%
x_{b}=x_{c}^{+}=\infty $ and $I=J=\left[ 0,\infty \right] .$

- If $\phi _{0}\left( b/2\right) =c+r\left( b/2\right) ^{2}\leq 0$, then $%
x_{b}<\infty $. Here $I=\left[ 0,x_{b}\right] $ and $J=\left[ 0,c\right]
\subseteq I$ provided $c\leq x_{b}.$

$\left( ii\right) $ (homographic map) $\phi _{0}\left( x\right) =c+rx/\left(
1+ax\right) $ with $x_{b}=\infty =x_{c}^{+}$. Here $I=\left[ 0,\infty
\right] $ and $J=\left[ c,c+r/a\right] \subset I.$ For this map, $g$ in (\ref
{e14}) is readily seen to be an homographic function itself, using the
matrix representation of an homography, matrix product translating into
composition of homographic maps.

$\left( iii\right) $ (Ricker map) $\phi _{0}\left( x\right) =c+rx\exp \left(
-ax\right) $ with $x_{b}=$ $x_{c}^{+}=\infty .$ Here $I=\left[ 0,\infty
\right] $ and $J=\left[ c,c+1/a\right] \subset I.$

$\left( iv\right) $ $\phi _{0}\left( x\right) =c+x_{*}-\sqrt{a\left(
x_{*}-x\right) ^{2}+b}$. Here $I=\left[ 0,x_{b}\right] $ and $J=\left[
0,c+x_{*}-\sqrt{b}\right] \subseteq I$ $.$

In these 4 examples, there is a smallest $\rho >0$ for which $\phi
_{0}\left( \rho \right) =\rho $.\newline

$\left( v\right) $ Consider the quadratic population model with immigration $%
\phi _{0}\left( x\right) =x+x^{2}+c$ with $\phi _{0}\left( 0\right) =c>0$.
In this example, $\phi _{0}\left( x\right) $ has no real fixed point and
clearly with $x_{n+1}=\phi _{0}\left( x_{n}\right) $, $x_{0}=x$, $x_{n}\sim
\lambda \left( x,c\right) ^{2^{n}}$, with $\lambda \left( x,c\right) >1$
depending on both $x$ and $c$. $x_{n}$ drifts to $\infty $ at doubly
exponential speed with $n$.

$\left( vi\right) $ Consider the quadratic population model with immigration 
$\phi _{0}\left( x\right) =x+c/\left( x+1\right) $ with $\phi _{0}\left(
0\right) =c>0$. In this example, $\phi _{0}\left( x\right) $ has no real
fixed point either and clearly with $x_{n+1}=\phi _{0}\left( x_{n}\right) $, 
$x_{0}=x$, $x_{n}\sim \sqrt{2cn}\rightarrow \infty $, whatever $x>0.$ $x_{n}$
drifts algebraically slowly to $\infty $.

\subsection{Logistic population models with or without immigration}

Consider the logistic population dynamical system without immigration 
\begin{equation}
x_{n+1}=rx_{n}\left( 1-x_{n}\right) ,x_{0}=x,  \label{e16}
\end{equation}
where $r\in \left( 0,4\right] .$ Let $y_{n}=ax_{n}+b$, with $b=\left(
1-a\right) /2.$ Then, if $a=r/R$%
\begin{equation}
y_{n+1}=c+Ry_{n}\left( 1-y_{n}\right) ,y_{0}=y,  \label{e17}
\end{equation}
where 
\begin{equation}
c=\frac{\left( r-1\right) ^{2}-\left( R-1\right) ^{2}}{4R}.  \label{e18}
\end{equation}
For the dynamics $y_{n}$ to be a logistic population dynamical system with
immigration, mapping some interval $I$ onto $J\subseteq I$, elementary
algebra shows that the pair $\left( r,R\right) $ must lie in one of the
shaded zone $Z$ of Figure $1$ below, where $\left| R\right| \leq 4$ and $%
r\in \left( 0,4\right] $.\newline

The detailed explanation of Figure $1$ is as follows: for any value of $R$
and $r$, there exists a $c$ such that the dynamics $rx(1-x)$ and $Ry(1-y)+c$
are related. For the latter to be an admissible population dynamics
including immigration, it has to satisfy several conditions. First, the
number $c$ has to be positive (which is related to the lines $R=r$ and $%
R=2-r $ on the graph). With this condition, the dynamics in $y$ has to
belong to one of three classes: $R>0$ (corresponding to the vertically
dashed area), $R<0$ and the dynamics never reaches zero (the diagonally
dashed area), or $R<0$ and the dynamics reaches zero (the horizontally
dashed area). The domain of the first-kind dynamics obviously corresponds to
the superior half-plane, while the domains for the second and the third one
are separated by $R=r(2-r)/2$. Another condition to be an admissible
dynamics is that it has to map one finite (because the $rx(1-x)$ is defined
on $\left[ 0,1\right] $) domain into itself. This condition is related to
the lines $R=r(r-2)(r^{2}-2r-4)/8$ for the first-kind dynamics, $R=-r$ for
the second-kind one and $r^{2}-R^{2}-2r-2R-4=0$ for the third-kind one.
Finally, to be consistent with the $rx(1-x)$ dynamics, the initial condition
in $y$ has to be mapped in an initial condition in $x$ that belongs to $%
\left[ 0,1\right] $, which is in relation to the line $r=R$.\newline

This means that, starting from a dynamical system of type (\ref{e17}) for
some $R$ chosen in the range $\left| R\right| \leq 4$, for any value of $r$
intersecting the shaded zone with the horizontal line of equation $y=R$,
there is a positive immigration rate $c$ given by (\ref{e17}) such that the
dynamics of the $y$s is mapped into the dynamics (\ref{e16}) using the
reverse affine transformation $x_{n}=\left( y_{n}-b\right) /a$. We conclude:%
\newline

\textbf{Proposition:} With $\left| R\right| \leq 4$, if $r\in \left(
0,4\right] $ is in the shaded zone $Z$ of the graph and also in the chaotic
range for (\ref{e16}), then the logistic population dynamical system (\ref
{e17}) with immigration rate $c$ given by (\ref{e18}) is chaotic.\newline

\begin{figure}[h]
\includegraphics[scale=0.7]{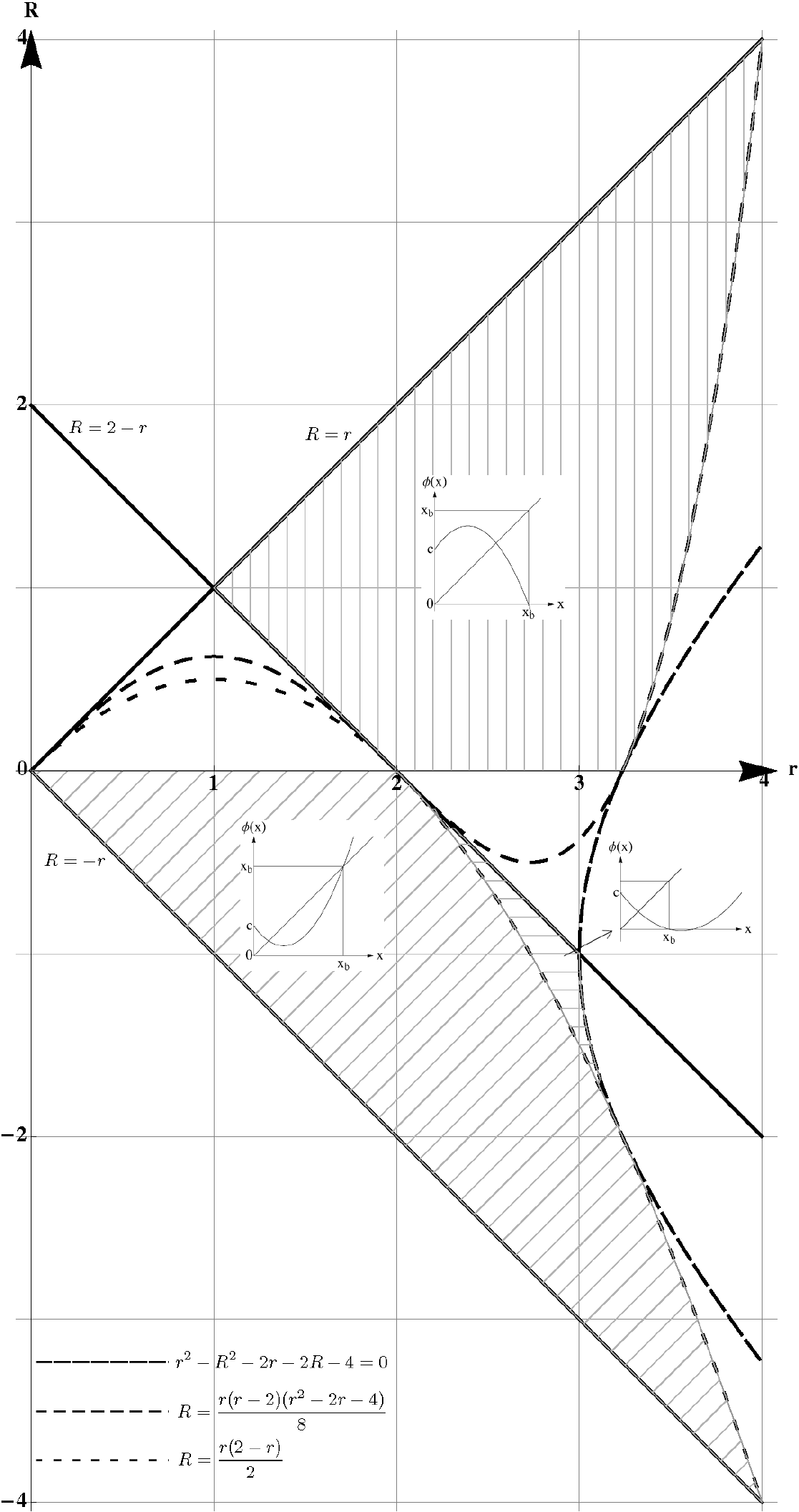}
\caption{Map of the admissible correspondences between $R$ and $r$. A point
in the dashed areas corresponds to an admissible transformation. The
miniatures of graphs indicate the kind of dynamics for $Ry(1-y)+c$. }
\end{figure}

Starting from a logistic model without immigration and with reproduction
rate $R>0$ (not) in the chaotic region, adding immigration $c$ can lead to a 
$\left( c,R\right) $ model (\ref{e17}) either (chaotic) or non chaotic,
depending on the chosen value of $r$: immigration can either (destabilize)
stabilize a (non) chaotic system, \cite{Sto}. We also note that when $R$ is
negative, but not too negative, there is no $r$ large enough to lie in the
chaotic region of (\ref{e16}). \newline

\textbf{Corollary:} With $\left| R\right| \leq 4$, suppose $r\in \left(
0,4\right] $ is in the shaded zone $Z$ of the graph. If $x_{n}=h^{-1}\left(
r^{n}h\left( x\right) \right) $ is the conjugacy representation (\ref{e7})
of $x_{n}$, then, with $a=r/R$ and $b=\left( 1-a\right) /2$ 
\begin{equation*}
y_{n}=ah^{-1}\left( r^{n}h\left( \left( y-b\right) /a\right) \right) +b
\end{equation*}
is a conjugacy solution $y_{n}$ of (\ref{e17}), started at $y$.\newline

\textbf{Corollary:} If (\ref{e17}) is chaotic and corresponding (\ref{e16})
has an invariant density $f$, then (\ref{e17}) has the image invariant
density: $\widetilde{f}\left( y\right) =f\left( \left( y-b\right) /a\right)
/a$.\newline

\emph{Example:} Let $R=-4$ and $r=4$ (in $Z$), so that $c=1$, corresponding
to the logistic population dynamical system with immigration 
\begin{equation}
y_{n+1}=1-4y_{n}\left( 1-y_{n}\right) ,y_{0}=y.  \label{e19}
\end{equation}

Here $a=-1$ and $b=1$. With $h\left( x\right) =\arcsin \left( \sqrt{x}%
\right) ^{2}$, we have 
\begin{equation*}
y_{n}=1-h^{-1}\left( 4^{n}h\left( \left( 1-y\right) \right) \right) ,
\end{equation*}
and this dynamics is chaotic.

It can be checked that the representation (\ref{e15}) takes the alternative
form 
\begin{equation*}
y_{n}=\sin ^{2}\left( \left( -2\right) ^{n}\arcsin \left( \sqrt{x}\right) +%
\frac{\pi }{6}\left( 1-\left( -2\right) ^{n}\right) \right) ,
\end{equation*}
corresponding to $g\left( x\right) =\arcsin \left( \sqrt{x}\right) $, $%
\overline{\lambda }=-2$ and $\rho =1/4$ (with $g\left( \rho \right) =\pi /6$%
), in the notations of subsection $4.1$.

(\ref{e16}) with $r=4$ has the invariant density $f\left( x\right) =\pi
^{-1}\left( x\left( 1-x\right) \right) ^{-1/2}$ and the logistic population
dynamical system with immigration (\ref{e19}) has the invariant density $%
\widetilde{f}\left( y\right) =f\left( \left( y-1\right) /\left( -1\right)
\right) =\pi ^{-1}\left( y\left( 1-y\right) \right) ^{-1/2}$, so identical
to $f$ (with Lyapounov exponent $\log 2$).

\section{Conclusion}

Using Carleman linearization techniques applied to $1-$dimensional
discrete-time population dynamical systems, we gave a technique to compute
the current population state $x_{n}$, for any initial point $x$, without
actually computing the intermediate values $x_{1},...,x_{n-1}$. This
technique was shown to be related to the characterization of the invariant
density measure, when it exists. But this is at the expense of the
computation of $h$\ and $h^{-1}$, which are ``simple'' special functions
only in some exceptional situations, such as for specific parameter values
of the logistic map with or without immigration. What ``simple'' means and
the class of models for which these functions are ``simple'' and/or lead to
chaotic behavior are largely open problems. This methodology has also
recently proved useful in the context of discrete-time branching process for
which the map $\phi $ is absolutely monotone: for the family of so-called
generalized linear-fractional branching processes first introduced in \cite
{Sag} and further studied in \cite{GH}, the function $h$ has a simple
structure and the iteration of $\phi $\ does not lead of course to chaos
being one-to-one on the unit interval.\newline

\textbf{Acknowledgments:}

T. Huillet acknowledges partial support from the ``Chaire \textit{%
Mod\'{e}lisation math\'{e}matique et biodiversit\'{e}''.} N. Grosjean and T.
Huillet also acknowledge support from the labex MME-DII Center of Excellence
(\textit{Mod\`{e}les math\'{e}matiques et \'{e}conomiques de la dynamique,
de l'incertitude et des interactions}, ANR-11-LABX-0023-01 project). The
authors would like to thank Genevi\`{e}ve Rollet for some fruitful
discussions.


\begin{thebibliography}{99}
\bibitem{Ber}  Berkolaiko, G.; Rabinovich, S.; Havlin, S. Analysis of
Carleman representation of analytical recursions. Journal of Mathematical
Analysis and Applications, vol. 224(1), 81-90, (1998).

\bibitem{Bowen}  Bowen, R. Invariant measures for Markov maps of the
interval. Comm. Math. Phys. Volume 69, no 1, 1-17, (1979).

\bibitem{Buni}  Bunimovich, L. A. A Transformation of the Circle. Math.
Notes. 8, no 2, 587-592, (1970).

\bibitem{Collet}  Collet, P.; Eckmann, J.P. Positive Liapunov exponents and
absolute continuity for maps of the interval. Ergodic Theory and Dynam.
Systems 3, 13-46, (1983).

\bibitem{Comtet}  Comtet, L. \emph{Analyse combinatoire.} Tomes 1 et 2.
Presses Universitaires de France, Paris, 1970.

\bibitem{Ding}  Ding, J.; Zhou, A. \emph{Statistical Properties of
Deterministic Systems. }Tsinghua University Texts and Springer, 2009.

\bibitem{Erdos}  Erd\"{o}s, P.; Jabotinsky, E. On analytic iteration.
Journal d'Analyse Math\'{e}matique. no 1, 361-376, (1960).

\bibitem{GH}  Grosjean, N.; Huillet, T. Additional aspects of the
generalized linear-fractional branching process. The Annals of the Institute
of Statistical Mathematics (to appear), online first, (2016).
arXiv:1607.01915.

\bibitem{GK}  Gralewicz, P.; Kowalski, K. Continuous time evolution from
iterated maps and Carleman linearization. Chaos, Solitons \& Fractals,
Volume 14, Issue 4, Pages 563-572, 2002.

\bibitem{Jab}  Jabotinsky, E. Analytic iteration. Transactions of the
American Mathematical Society 108 (3): 457-477, (1963).

\bibitem{Jako}  Jakobson, M. V. Absolutely Continuous Invariant Measures for
one-parameter Families of One-dimensional Maps. Commun. Math. Phys., 81,
39-88, (1981).

\bibitem{Ko}  Kowalski, K.; Steeb, W-H. \emph{Nonlinear dynamical systems
and Carleman linearization.} World Scientific Publishing Co. Pte. Ltd.
Singapore, 1991.

\bibitem{Las}  Lasota A.; Mackey, M. C. \emph{Chaos, Fractals, and Noise.
Stochastic Aspects of Dynamics. }Applied Mathematical Sciences, Volume 97,
Springer-Verlag, New York Inc, 1994.

\bibitem{Luev}  Lu\'{e}vano, J-R. ; Pina, E. The Schr\"{o}der functional
equation and its relation to the invariant measures of chaotic maps. J.
Phys. A: Math. Theor. 41, 265101, (2008).

\bibitem{Rab}  Rabinovich, S.; Berkolaiko, G.; Havlin, S. Solving nonlinear
recursions. Jal. Math. Phys. 37(11), (1996).

\bibitem{Ruelle}  Ruelle, D. Applications conservant une mesure absolument
continue par rapport \`{a} $dx$ sur $[0,1]$. Commun. Math. Phys. 55, 47--51,
(1977).

\bibitem{Sag}  Sagitov S.; Lindo A. A special family of Galton-Watson
processes with explosions. In Branching Processes and Their Applications.\
Lecture Notes in Statistics - Proceedings. (I.M. del Puerto et al eds.)
Springer, Berlin, 2016 (to appear). arxiv.org/pdf/1502.07538, (2015).

\bibitem{Sto}  Stone, L.; Hart, D. Effects of immigration on the dynamics of
simple population models. Theoretical Population Biology 55, 227-234, (1999).

\bibitem{Thun}  Thunberg, H. Periodicity versus chaos in one-dimensional
dynamics. SIAM Reviews 43, No. 1, 330, (2001).
\end{thebibliography}
\end{document}